\documentclass[twocolumn,english]{revtex4}
\usepackage[T1]{fontenc}
\usepackage[latin9]{inputenc}
\usepackage{amsmath}
\usepackage{amssymb}
\usepackage{graphicx}

\makeatletter
\@ifundefined{textcolor}{}
{%
 \definecolor{BLACK}{gray}{0}
 \definecolor{WHITE}{gray}{1}
 \definecolor{RED}{rgb}{1,0,0}
 \definecolor{GREEN}{rgb}{0,1,0}
 \definecolor{BLUE}{rgb}{0,0,1}
 \definecolor{CYAN}{cmyk}{1,0,0,0}
 \definecolor{MAGENTA}{cmyk}{0,1,0,0}
 \definecolor{YELLOW}{cmyk}{0,0,1,0}
}

\usepackage{babel}

\makeatother

\usepackage{babel}
\begin{document}

\title{Pattern formation in systems with multiple delayed feedbacks }

\author{Serhiy Yanchuk$^{1}$ and Giovanni Giacomelli$^{2}$}

\affiliation{$^{1}$Institute of Mathematics, Humboldt University of Berlin, Unter
den Linden 6, 10099 Berlin, Germany\\
 $^{2}$CNR - Istituto dei Sistemi Complessi - via Madonna del Piano
10, I-50019 Sesto Fiorentino (FI), Italy}
\begin{abstract}
Dynamical systems with complex delayed interactions arise commonly
when propagation times are significant, yielding complicated oscillatory
instabilities. In this Letter, we introduce a class of systems with
multiple, hierarchically long time delays, and using a suitable space-time
representation we uncover features otherwise hidden in their temporal
dynamics. The behaviour in the case of two delays is shown to ''encode''
two-dimensional spiral defects and defects turbulence. A multiple
scale analysis sets the equivalence to a complex Ginzburg-Landau equation,
and a novel criterium for the attainment of the long-delay regime
is introduced. We also demonstrate this phenomenon for a semiconductor
laser with two delayed optical feedbacks. 
\end{abstract}


\maketitle
Systems with time delays are common in many fields, ranging from optics
(e.g. laser with feedback \cite{Li2011,Zamora-Munt2010,Giacomelli2012,Soriano2013}),
vehicle systems \cite{Szalai2013}, to neural networks \cite{Izhikevich2006},
information processing \cite{Appeltant2011}, and many others \cite{Erneux2009}.
A finite propagation velocity of the information introduces in such
systems a new relevant scale, which is comparable or higher than the
intrinsic timescales. It has been shown that the complexity of such
systems, e.g. the dimension of attractors, is finite and it grows
linearly with time delay \cite{Farmer1982}; moreover, the spectrum
of Lyapunov exponents approaches a continuous limit for long delay
\cite{Giacomelli1995,Heiligenthal2011,DHuys2013}. As a result, in
this case essentially high-dimensional phenomena can occur such as
spatio-temporal chaos \cite{Giacomelli1996}, square waves \cite{Erneux2009},
Eckhaus destabilization \cite{Wolfrum2006}, or coarsening \cite{Giacomelli2012}.
In the above mentioned situations, the system involves one long delay,
which can be interpreted as the size of a one-dimensional, spatially
extended system \cite{Arecchi1992,Giacomelli1996}. This approach
has proven to be instrumental in explaining new phenomena in systems
with time delays \cite{Giacomelli1994,Larger2013}.

In this Letter, we show that many new challenging problems arise when
a system is subject to several delayed feedbacks acting on different
scales. In contrast to the single delay situation, essentially new
phenomena occur, related to higher spatial dimensions involved in
the dynamics, such as spirals or defect turbulence. As an illustration,
we consider a specific physical system, namely, a model of a semiconductor
laser with two optical feedbacks.

A simple paradigmatic setup for the multiple delays case is the following
system 
\begin{equation}
\dot{z}=az+bz_{\tau_{1}}+cz_{\tau_{2}}+dz|z|^{2}.\label{eq:SL}
\end{equation}
Eq.~(\ref{eq:SL}) describes a very general situation: the interplay
of the oscillatory instability (Hopf bifurcation) and two delayed
feedbacks $z_{\tau_{i}}=z(t-\tau_{i})$, that we consider acting on
different timescales $1\ll\tau_{1}\ll\tau_{2}$. The variable $z(t)$
is complex, and the parameters $a$, $b$, and $c$ determine the
instantaneous, $\tau_{1}$-, and $\tau_{2}$-feedback rates, respectively.
The instantaneous part of the system (without feedback) is known as
the normal form for the Hopf bifurcation.

The following basic questions arise: what kind of new phenomena can
be observed in systems with several delayed feedbacks? Can one relate
the dynamics of such systems to spatially extended systems with several
spatial dimensions? In the case of positive answer, under which conditions?
Is it possible to observe such essentially 2D phenomena as, e.g.,
spiral waves in purely temporal delay systems (\ref{eq:SL}), which
obey the causality principle with respect to the time? In this Letter
we address the above questions. In particular, we show that such inherently
2D patterns as spiral defects or defect turbulence \cite{Chate1996},
are typical behaviors of system (\ref{eq:SL}). Moreover, they can
be generically found in a semiconductor laser model with two optical
feedbacks.

\begin{figure}
\includegraphics[width=1\columnwidth]{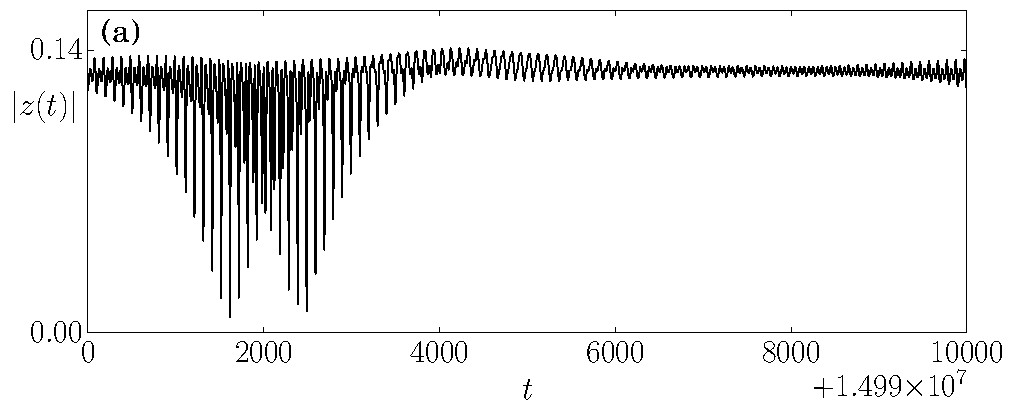}

\includegraphics[width=1\columnwidth]{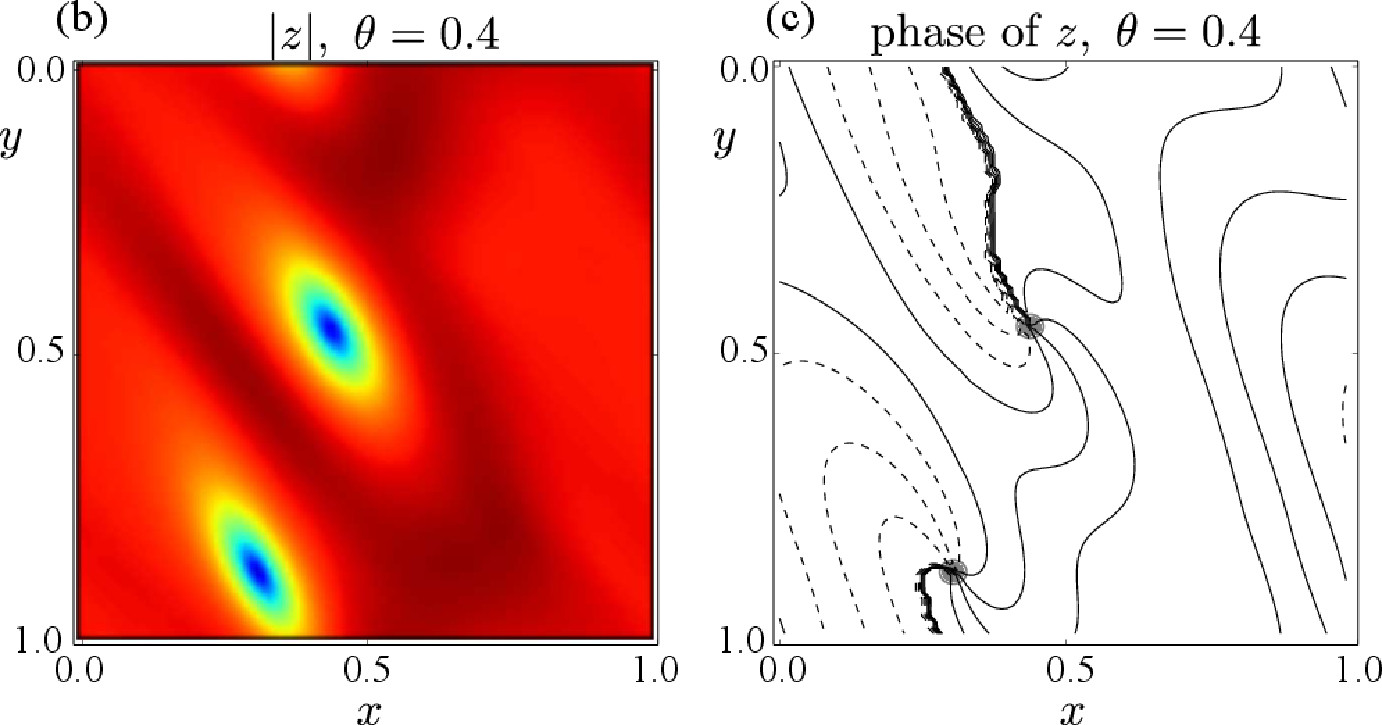}

\caption{\label{fig:FS} Spiral defects in system with delays (\ref{eq:SL}).
(a) Typical time series of the absolute value $|z(t)|$. Spatio-temporal
representation of the time series using pseudo-space coordinates (\ref{eq:Space})
reveals the spiral defects: (b) Snapshot of the spatial profile in
the pseudo-space coordinates $(x,y)$ for $\theta_{0}=0.4$. (c) Constant
level lines for the phase of $z$. Circles denote the positions of
defects. Parameters: $a=-0.985$, $b=0.4$, $c=0.6$ (corresponding
to $P=0.015$), $d=-0.75+i$, $\tau_{1}=100$, and $\tau_{2}=10000$.
Initial conditions are chosen randomly. }
\end{figure}

\begin{figure}
\includegraphics[width=1\columnwidth]{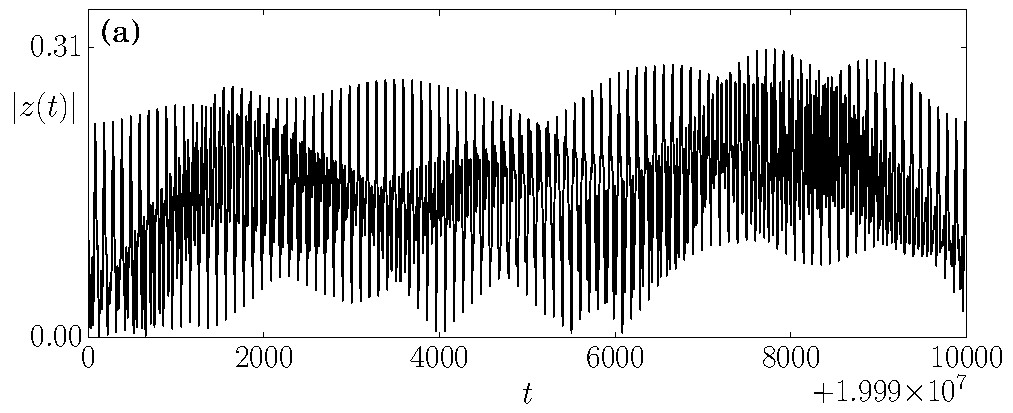}

\includegraphics[width=1\columnwidth]{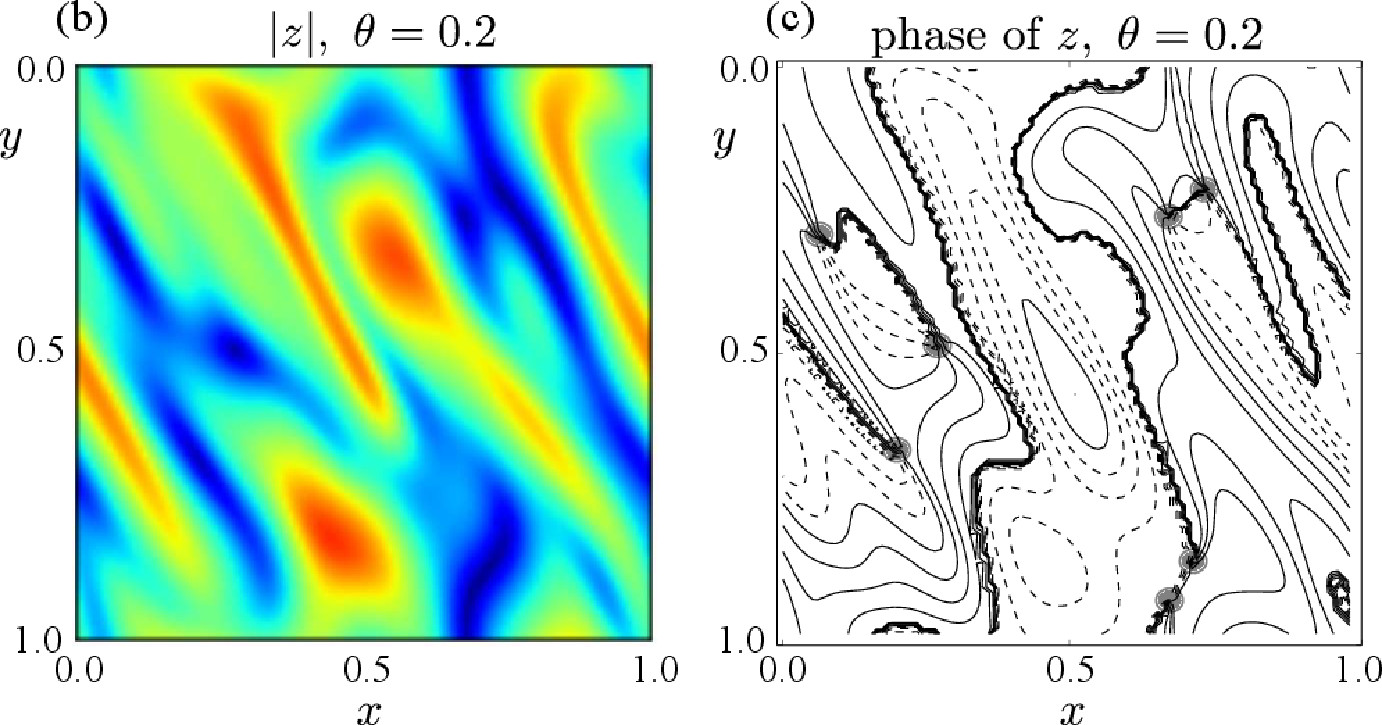}

\caption{\label{fig:DT} Defects turbulence in delayed system (\ref{eq:SL}).
Same as in Fig.~\ref{fig:FS} for different value of $d=-0.1+i$.
Spatio-temporal representation in (b) and (c) reveals defects turbulence.}
\end{figure}

We start with numerical examples. Figures~\ref{fig:FS} and \ref{fig:DT}
show solutions of Eq.~(\ref{eq:SL}) for two different parameter
choices. Time series in Figs.~\ref{fig:FS}(a) and \ref{fig:DT}(a)
exhibit oscillations on different timescales related approximately
to the delay times. However, an appropriate spatio-temporal representation
of the data {[}see e.g. Figs.~\ref{fig:FS}(b-c) and \ref{fig:DT}(b-c){]}
reveals clearly the nature of the dynamical behaviors. More details
on the appropriate spatio-temporal representation of these purely
temporal data will be given later, but one can readily observe that
the first case corresponds to a (frozen) spiral (FS) defects solution,
see Fig.~\ref{fig:FS}(b,c). The positions of the two coexisting
spiral defects are shown by the dots, where the level lines for the
phase meet. Consequently, the phase is not defined there and $|z|=0$.
The solution shown in Fig.~\ref{fig:DT} corresponds instead to the
defect turbulence (DT) regime. One can observe that the modulation
of the amplitude $\left|z(t)\right|$ starts to approach the zero
level in a random-like manner. In this case, the corresponding spatial
representation (see Fig.~\ref{fig:DT}(b,c)) reveals DT, i.e. the
non-regular motions of the spiral defects. The plots correspond to
snapshots in time.

In the following, we explain why the observed behaviors are typical
and show how to relate the dynamics of (\ref{eq:SL}) to the complex
Ginzburg-Landau equation on a 2D spatial domain. In particular, we
show that the function $z(t)$ on the time interval of the length
$\tau_{2}$ corresponds to a snapshot of a 2D \emph{spatial} function
$\Phi(x,y)$. The corresponding pseudo-spatial coordinates $x$ and
$y$ introduced later by Eq.~(\ref{eq:Space}) are different scales
of the time. We will show, that the parameters of (\ref{eq:SL}) leading
to the FS (resp. DT) can be mapped uniquely to the parameters of the
Ginzburg-Landau system, for which the same phenomenologies are observed
\cite{Chate1996}. This behavior is observed robustly for all tested
random initial conditions for an interval of parameters.

\emph{Normal form equation.} The long time delay $\tau_{1}$ can be
written as $\tau_{1}=1/\varepsilon$ with a small positive parameter
$\varepsilon$, and $\tau_{2}=\kappa/\varepsilon^{2}$ with some positive
$\kappa$. With such notations, the scale separation $1\ll\tau_{1}\ll\tau_{2}$
is satisfied. Notice that this also gives an indication how one should
proceed in the case of more than two delays.

In order to derive a normal form describing universally the dynamics
close to the destabilization of system (\ref{eq:SL}), the multiple
scale ansatz $z(t):=\varepsilon u\left(\varepsilon t,\varepsilon^{2}t,\varepsilon^{3}t,\varepsilon^{4}t\right)$
is used. More precisely, substituting this ansatz as well as the perturbation
parameter $\varepsilon^{2}p=a+|b|+|c|$ in (\ref{eq:SL}), and time
delays $\tau_{1}=1/\varepsilon$, $\tau_{2}=\kappa/\varepsilon^{2}$,
one obtains several separate solvability conditions for different
orders of $\varepsilon$. The resulting equation is the Ginzburg-Landau
partial differential equation 
\begin{equation}
\Phi_{\theta}=p\Phi+a_{1}\Phi_{x}+a_{2}\Phi_{y}+a_{3}\Phi_{xx}+a_{4}\Phi_{xy}+a_{5}\Phi_{yy}+d\Phi|\Phi|^{2}\label{eq:GL}
\end{equation}
for a function $\Phi(\theta,x,y)$, which is related to the solutions
of (\ref{eq:SL}) by $z(t)=\varepsilon\Phi\left(\theta,x,y\right)$,
where 
\begin{equation}
\theta=\varepsilon^{4}\delta t,\quad x=\varepsilon t\left(1-\delta\varepsilon^{2}\right),\quad y=\varepsilon^{2}t\left(1-\left|b\right|\delta\varepsilon\right),\label{eq:Space}
\end{equation}
and $\delta=-\left(a+\left|b\right|\right)^{-1}>0$. The new spatial
variables $x$ and $y$ are different timescales of the original time
$t$, and the new time variable $\theta$ is the slow time scale $\varepsilon^{4}t$.
Therefore, the new spatial and temporal variables can be called pseudo-space
and pseudo-time. The coefficients in (\ref{eq:GL}) are $a_{1}=a_{4}=\delta|b|$,
$a_{2}=-1+\delta\left|b\right|^{2}$, $a_{3}=\delta/2$, and $a_{5}=-\delta a\left|b\right|/2$.
One can note that the diffusion coefficients in this equation are
real. The dynamics of (\ref{eq:GL}) is known \cite{Chate1996,Aranson2002}
to possess various phase transitions, FS (e.g. for $d=-0.75+i$),
and DT (e.g. for $d=-0.1+i$). We found a good correspondence between
the dynamics of systems (\ref{eq:GL}) and (\ref{eq:SL}), taking
into account the relation (\ref{eq:Space}) between them. Although
a systematic parametric investigation is out of the scope of this
Letter, the examples of FS and DT for the above mentioned parameter
values shown in Figs.~\ref{fig:FS} and \ref{fig:DT} are well reproduced.
Moreover, the observed dynamics is robust with respect to small variations
of parameters. We remark that the observed phenomena are not possible
in systems with one time delay, since they arise from the two-dimensional
space $(x,y)$ of the normal form equation.

\emph{Drift and comoving Lyapunov exponents.} The spatial coordinates
in (\ref{eq:Space}) can be rewritten as $x=\bar{x}-\delta\bar{u}$
and $y=\bar{y}-|b|\delta\bar{u}$, where $\bar{x}=\varepsilon t$,
$\bar{y}=\varepsilon^{2}t$, and $\bar{u}=\varepsilon^{3}t$. As a
consequence, we can infer the existence of a (fast) drift along the
vector $\mathbf{V}_{d}=(-1,-|b|)$ in the ``naive'' coordinates
$(\bar{x},\bar{y})$. The corrected coordinates (\ref{eq:Space})
eliminate this drift so that the remaining variables are governed
by the Ginzburg-Landau equation (\ref{eq:GL}).

The above phenomenon is a consequence of the properties of the maximal
comoving (or convective) Lyapunov exponent $\Lambda$ \cite{Deissler1987}.
In the spherical coordinates $\bar{u}=\rho\cos\alpha$, $\bar{y}=\rho\sin\alpha\cos\beta$,
$\bar{x}=\rho\sin\alpha\sin\beta$, it is found that 
\begin{multline}
\Lambda(\alpha,\beta)=a\sin\alpha\sin\beta+\big(1+\log{(|b|\tan\beta)}\big)\sin\alpha\cos\beta\\
+\big(1+\log{(|c|\sin\beta\tan\alpha)}\big)\cos\alpha.\label{MCLE}
\end{multline}
Details of the calculation will be presented elsewhere. A geometrical
interpretation can be introduced using the velocity $\mathbf{V}=(\sin\beta\tan\alpha,\cos\beta\tan\alpha)$,
along which the perturbations evolve with a multiplier $e^{\Lambda(\alpha,\beta)}$.
The propagation cone's boundaries can be defined as the set $(\alpha,\beta)$
such that $\Lambda(\alpha,\beta)=0$. The bifurcation point, attained
when the maximum of $\Lambda$ is equal to zero, is obtained at $\mathbf{V}=\delta\mathbf{V}_{d}$,
corresponding to $(\alpha_{0},\beta_{0})=(\tan^{-1}(-\delta\sqrt{1+|b|^{2}}),\tan^{-1}(|b|^{-1}))$.
Note that the direction $\mathbf{V}_{d}$ is also given by the multiscale
method above. The above result extends the standard linear stability
analysis by indicating the direction along which the destabilization
takes place. We notice that the comoving exponent diverges logarithmically
close to the axis $\alpha=0$ and $\beta=0$, i.e. instantaneous propagations
are forbidden. In the opposite limit, $\alpha\to\pi/2$ (resp. $\beta\to\pi/2$),
$\Lambda$ approaches the value for the single delay case $c=0$ ($b=0$).
Finally when both $\alpha,\beta\to\pi/2$ (infinite velocity), $\Lambda=a$
and the dynamics is governed by the local term as expected.

\begin{figure}[t]
\vskip 4mm \includegraphics[width=0.48\linewidth]{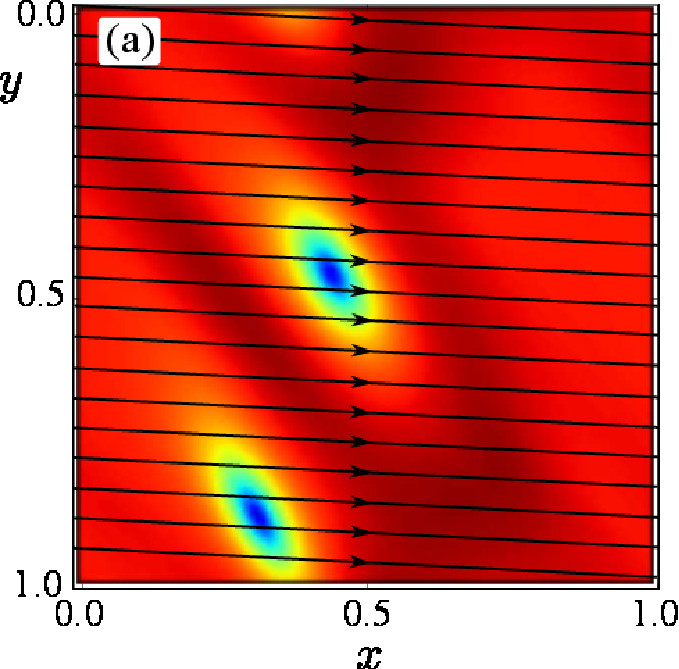}\includegraphics[width=0.51\linewidth]{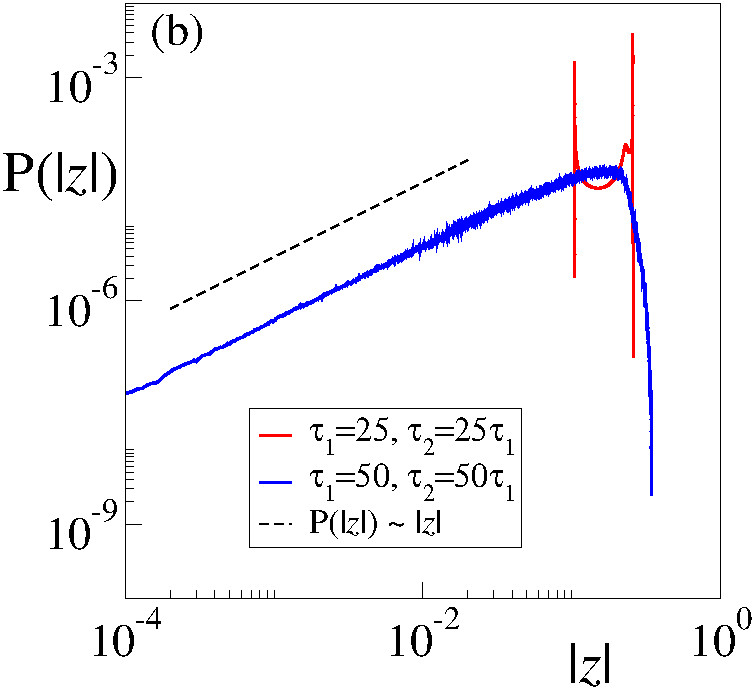}
\caption{\label{fig:SmallT}\textquotedbl{}Small\textquotedbl{} delays effect.
(a) One-parameter curve $x(t),y(t)$ in the pseudo-space determined
by (\ref{eq:Space}) for $\varepsilon=0.05$. For larger distances
between the branches (smaller delays), the line does not resolve the
cores of the spiral of the corresponding GL model. (b) numerical histograms
of $|z|$ for the DT regime (parameters as in Fig.~\ref{fig:DT})
for increasing delays values. Histograms for smaller delays (here,
$\tau_{1}=25$, $\tau_{2}=25\tau_{1}$) correspond to bounded, periodic
solutions with no defects, reached after a transient. A tail in the
distribution appears for highers delays (here, $\tau_{1}=50,$ $\tau_{2}=50\tau_{1}$).
The dashed line is a reference curve $P(|z|)\sim|z|^{1}$. }
\end{figure}

\emph{On long delay approximation.} Concerning the relation between
the delay system (\ref{eq:SL}) and the normal form (\ref{eq:GL}),
the following questions arise: to what extent the equivalence is founded?
Under which conditions the delays are large \textquotedbl{}enough\textquotedbl{}?
Dynamically, the absence of the anomalous Lyapunov exponents \cite{Giacomelli1995}
is required, or, equivalently, the absence of the strong chaos \cite{Heiligenthal2011}.
Numerically, with the decreasing of delays, the spatio-temporal structures
become transients towards a periodic or constant amplitude ($|z|=\mbox{const}$)
state. As a matter of fact, a solution of the delay system evolves
along the one-parametric line $(\theta(t),x(t),y(t))$ defined by
(\ref{eq:Space}) in the pseudo space $(\theta,x,y)$, see Fig.~\ref{fig:SmallT}(a).
In order to have a good correspondence between the solutions of delay
system (\ref{eq:SL}) and the normal form through the parametrization
(\ref{eq:Space}), the line $(x(t),x(t))$ should wind up the space
$(x,y)$ sufficiently densely. In the leading order, this line satisfies
$y\approx\varepsilon x$ and it is wrapped periodically at $x=0$
and $x=1$, see Fig.~\ref{fig:SmallT}(a). The distance between the
neighboring branches is $\sim\varepsilon$ which determines the ``discretization''
level. Thus, high delays imply dense covering of the (pseudo) space
plane, as expected in the thermodynamic limit. However, when such
density is too small the dynamics changes drastically and the delay
system behaves quite differently from the corresponding normal form.

To illustrate such a behavior, we present in Fig.~\ref{fig:SmallT}(b)
the analysis of the amplitude $|z|$ statistics in the defect turbulence
regime for the model (\ref{eq:SL}). For small delays the dynamics
relaxes to a stationary oscillating regime after a transient, with
the corresponding histogram showing a shape very close to that obtained
from a sinusoidal signal. For higher $\tau$'s , the histogram start
displaying a power-law tail $P(|z|)\sim|z|^{1}$ for $|z|\to0$, indicating
the stable appearance of defects and the attainment of the long-delayed
regime.

The scaling exponent can be obtained analytically for an arbitrary
number of delays and equations. In the DT regime, defects are the
spatial points where $|\Phi|=0$ in a $N$-dimensional space ($N$
being the number of delays) and form a set $D$ that we can assume
of constant density in space. In our case of $N=2$ these are point
defects, for $N=3$ line defects, etc.. In general, it holds that
$\mathrm{codim}(D)=2$ in the $N$-dimensional space $\{x_{1},x_{2},..x_{N}\}$
where $\{x_{i}\}$ are the pseudo-spatial coordinates. The delay equation(s)
dynamics approaches $D$ along the domain line $L=\{x_{1}(t),x_{2}(t),..x_{N}(t):t\in\mathbb{R}\}$.
The vicinity of defects in the pseudo space affects the amplitude
statistics of the delay dynamics, which will be depending only on
$\mathrm{codim}(D)=2$ and on $\mathrm{dim}(L)=1$. Thus, the scaling
exponent does not depend on $N$ or on the number of equations and
it can be shown to be equal to $1$.

\emph{Semiconductor laser with two optical feedbacks.} The results
obtained from the study of the normal form (\ref{eq:SL}) are expected
to apply to a wide class of physical systems. In the following, we
consider a Lang-Kobayashi-type model \cite{Lang1980} of a single
mode semiconductor laser with optical feedback, generalized to a double
external cavity configuration: 
\begin{equation}
\begin{aligned} & E'(t)=(1+i\alpha)n(t)E(t)+\eta_{1}E(t-\tau_{1})+\eta_{2}E(t-\tau_{2}),\\
 & Tn'(t)=J-n(t)-(2n(t)+1)|E(t)|^{2}.
\end{aligned}
\label{eq:LK}
\end{equation}
$E(t)$ is the complex electric field and $n(t)$ the excess carrier
density. The system parameters are the excess pump current $J$, the
external cavities round trip times $\tau_{1}$ and $\tau_{2}$ measured
in units of the photon lifetime, and the feedback strengths $\eta_{1}$
and $\eta_{2}$. The linewidth enhancement factor $\alpha$ is specific
for semiconductor lasers and affects many aspects of their behavior
(see e.g. \cite{Soriano2013}). We present here two examples of the
dynamics of (\ref{eq:LK}) in the case of $\alpha=2$ and $\alpha=4$.
Suitable laser devices can be employed to realize the corresponding
experiments; in fact, such range is typical and e.g. measurements
in-between have been reported \cite{Barland2005}.

\begin{figure}
\vskip 1mm \includegraphics[width=0.8\linewidth]{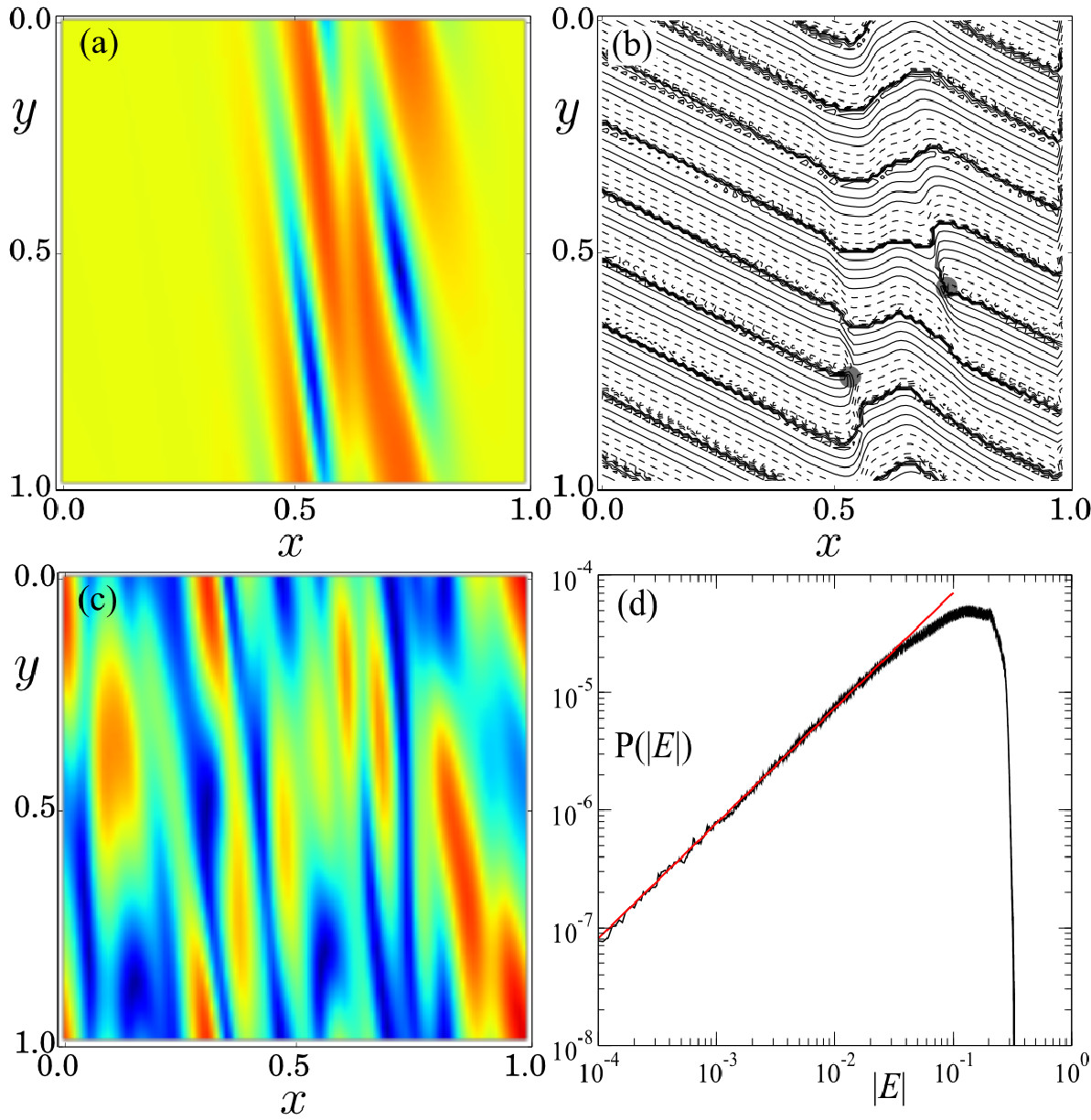} \caption{\label{fig:LK} Dynamics of the solution $E$ of the system (\ref{eq:LK}),
represented as snapshot in the pseudo space, for the parameter values:
$\tau_{1}=10^{2}$, $\tau_{2}=10^{4}$, $\eta_{1}=\eta_{2}=0.1$,
$T=10^{2}$, and $J=-0.17$. (a), (b): amplitude and phase of $E$
for $\alpha=2$, showing the occurrence of spiral defects. (c): amplitude
of $E$, defects turbulence regime for $\alpha=4$. (d) Statistics
of the field amplitude in the case (c); the line is a power-law fit
of the tail with exponent $0.98$.}
\end{figure}

In our case, shortly after the destabilization of the ''off-state''
$E=0$, a multifrequency oscillating behavior is found, corresponding
to FS ($\alpha=2$, Fig.~\ref{fig:LK}(a-b)) or DT ($\alpha=4$,
Fig.~\ref{fig:LK}(c)). These regimes are very similar to those shown
in Fig.~\ref{fig:FS} and Fig.~\ref{fig:DT} respectively. In order
to compare their statistical properties, we report also the distribution
of the field amplitude $|E|$ (Fig.~\ref{fig:LK}(d)). Its shape
is indeed consistent with the previous results and the scaling of
the tail marks the sign of the long-delay regime as well. We point
out also how $\alpha$ appears to be an effective parameter switching
between just drifting defects {[}Figs.~\ref{fig:LK}(a,b){]} and
irregularly moving defects {[}Fig.~\ref{fig:LK}(c){]}, thus suggesting
which kind of behavior could be expected for different laser devices.

In conclusion, we have discussed a class of systems describing the
interplay of the oscillatory instability with multiple, hierarchically
long, delayed feedbacks. We have shown that a generalized spatio-temporal
representation is able to uncover multiscale features otherwise hidden
in the complex temporal dynamics. In the case of two delays, the existence
of regimes of FS and DT has been evidenced. By means of a multiple
scale analysis, an equivalence is shown to a two-dimensional Complex-Ginzburg
Landau equation. The attainment of the long-delay regime have been
also analyzed. Finally, we showed how the above phenomena occur in
the case study of a semiconductor laser with two external cavity optical
feedbacks, a generalization of a well known and studied configuration.
As a perspective, our approach can be applied in several experimental
setups and in the study of higher-dimensional pattern formations in
delay systems, such that the existence and characterization of line
defects in the three delays case. Moreover, we expect that this formalism
could be generalized for other types of bifurcations as well and applied
to the study of specific experimental systems, such as delayed networks
like those commonly found in optical communications.

We acknowledge the DFG for financial support in the framework of International
Research Training Group 1740 and useful discussions with A. Politi.

\bibliographystyle{apsrev}

\section{Appendix: Destabilization of the steady state}

Prior to deriving the Ginzburg-Landau normal form, it is important
to study the destabilization of the steady state $z=0$. The type
of the destabilization will give us the key to what kind of normal
form is governing the dynamics.

The characteristic equation, which determines the stability of the
zero steady state $z=0$ is obtained by linearizing Eq.~(1, main
text) and substituting $z=e^{\lambda t}$: 
\begin{equation}
\lambda-a-be^{-\lambda/\varepsilon}-ce^{-\lambda\kappa/\varepsilon^{2}}=0.\label{eq:CHE}
\end{equation}
Stability of the steady state is equivalent to that all roots $\lambda$
of (\ref{eq:CHE}) have negative real parts. Although the solutions
to (\ref{eq:CHE}) are not given explicitly, their approximations
can be found using the smallness of $\varepsilon$ \cite{DHuys2013,Heiligenthal2011,Wolfrum2010}
(largeness of the delays) 
\begin{equation}
\lambda=\gamma_{0}+i\omega_{0}+\varepsilon\left(\gamma_{1}+i\omega_{1}\right)+\varepsilon^{2}\left(\gamma_{2}+i\omega_{2}\right),\label{eq:MS}
\end{equation}
where $\gamma_{j}$ and $\omega_{j}$ are real. Depending on the leading
terms in the real part of this expansion, the system may develop different
types of instabilities: if $\gamma_{0}>0$, there appear strong instability
induced by the instantaneous term \cite{Lepri1993,Wolfrum2006,Yanchuk2010a,Heiligenthal2011,Lichtner2011}.
If $\gamma_{0}=0$ but $\gamma_{1}>0$, there appears a weak instability
by the effect of the $\tau_{1}$-feedback. In this case, the $\tau_{2}$-feedback
does not play any role. Hence, in order for the second delay to play
the destabilizing role, one needs $\gamma_{0}=\gamma_{1}=0$ and $\gamma_{2}$
becoming positive. By requiring this and substituting (\ref{eq:MS})
into (\ref{eq:CHE}), one can arrive to the following conditions for
the parameters of the system: $a<0$ and $|b|<|a|$. Moreover, the
leading terms in the real part of $\lambda$ can be found explicitly
in this case 
\[
\gamma_{2}(\omega_{0},\phi)=-\frac{1}{2\kappa}\ln\frac{\left(a+|b|\cos\phi\right)^{2}+\left(\omega_{0}+|b|\sin\phi\right)^{2}}{|c|^{2}},
\]
where $\phi=-\omega_{1}-\frac{\omega_{0}}{\varepsilon}+\arg(b)$.
If the condition $|c|<-a-|b|$ is satisfied, the function $\gamma_{2}$
is negative for all $\omega_{0}$ and $\phi$, implying the stability
of the steady state. Otherwise, $\gamma_{2}$ becomes positive and
the steady state is unstable for all small enough $\varepsilon$.
In this case, a nontrivial dynamics is expected.

The obtained conditions determine when the $\tau_{2}$-feedback destabilize
the steady state. Namely, we have $a<0$, $|b|<|a|$, and $P=a+|b|+|c|$,
with $P$ as the destabilization parameter. The desired destabilization
occurs for positive values of $P$. For our purposes, the destabilization
parameter $P$ is chosen as $P=\varepsilon^{2}p=a+|b|+|c|$, where
the choice of the smallness factor of $\varepsilon^{2}p$ prevents
the unbounded increasing of the number of unstable linear modes (unstable
solutions of (\ref{eq:CHE})) with the decreasing of $\varepsilon$
{[}see more details in \cite{Wolfrum2006}{]}.

\onecolumngrid

\section{Appendix: Normal form equation}

Here we present a sketch of the formal derivation of the normal form
equation (2) from the main text as well as its boundary conditions.
System with time delay close to the destabilization has the following
perturbative form:

\begin{equation}
z'(t)=\left(a+p_{a}\varepsilon^{2}\right)z(t)+\left(Be^{i\phi}+p_{b}\varepsilon^{2}\right)z\left(t-\frac{1}{\varepsilon}\right)+\left(\left(-a-B\right)e^{i\xi}+p_{c}\varepsilon^{2}\right)z\left(t-\frac{\kappa}{\varepsilon^{2}}\right)-dz(t)|z(t)|^{2}\label{eq:perturbative}
\end{equation}
 where $a<0,$ $B<-a.$ As follows from the spectrum analysis (previous
section), the destabilization takes place at $p_{a}=p_{b}=p_{c}=0$.
Our aim is to obtain an equivalent amplitude equations. For this,
the multiscale ansatz $z(t)=\varepsilon\left[u_{1}\left(T_{1},T_{2},T_{3},T_{4},\dots\right)+\varepsilon u_{2}\left(\cdots\right)+\varepsilon^{2}u_{3}\left(\cdots\right)+\dots\right],$
$T_{j}=\varepsilon^{j}t$, is substituted in (\ref{eq:perturbative})
and the obtained expression is expanded in powers of the small parameter
$\varepsilon$. Afterward, terms with the same smallness factor $\varepsilon^{j}$
are compared.

In particular, for $\varepsilon^{1}$ we obtain the following solvability
conditions 
\[
u_{1}(T_{1},T_{2},\dots)=e^{i\phi}u_{1}(T_{1}-1,T_{2},\dots)
\]
and 
\[
u_{1}(T_{1},T_{2},\dots)=e^{i\xi}u_{1}(T_{1}-\frac{\kappa}{\varepsilon},T_{2}-\kappa,\dots)=e^{i\left(\xi-\phi\left[\frac{\kappa}{\varepsilon}\right]_{in}\right)}u_{1}(T_{1}-\left\{ \frac{\kappa}{\varepsilon}\right\} _{f},T_{2}-\kappa,\dots),
\]
where $\left\{ \cdot\right\} _{f}$ is the fractional and $\left[\cdot\right]_{in}$
integer part of a number. These conditions will result into the boundary
conditions of the normal form equation.

For $\varepsilon^{2}$, we obtain 
\[
\partial_{T_{1}}u_{1}=-B\partial_{T_{2}}u_{1}+\left(a+B\right)\partial_{T_{3}}u_{1}.
\]
This condition connects $\partial_{T_{1}}$ with $\partial_{T_{2}}$
and $\partial_{T_{3}}$ (a kind of transport). This means that the
solution depends only on 3 variables: 
\[
x=T_{1}+\frac{1}{a+B}T_{3};\quad y=T_{2}+\frac{B}{a+B}T_{3};\quad\theta=T_{4}.
\]
Hence, instead of $u_{1}$, we introduce new function $\Phi$ as follows
\begin{equation}
u_{1}(T_{1},T_{2},T_{3},T_{4})=\Phi\left(T_{1}+\frac{1}{a+B}T_{3},T_{2}+\frac{B}{a+B}T_{3},T_{4}\right)=:\Phi(x,y,\theta).\label{phi}
\end{equation}

Finally, $\varepsilon^{3}$ terms lead to \textbf{ 
\[
-\left(a+B\right)\partial_{T_{4}}u_{1}=Pu_{1}-\partial_{T_{2}}u_{1}-B\partial_{T_{3}}u_{1}+\frac{1}{2}B\partial_{T_{2}T_{2}}u_{1}-\left(a+B\right)\frac{1}{2}\partial_{T_{3}T_{3}}u_{1}-du_{1}|u_{1}|^{2},
\]
}where $P=p_{a}+p_{b}e^{-i\phi}+p_{c}e^{-i\xi}$. Rewriting the obtained
equation for the new function $\Phi$, we obtain the Ginzburg-Landau
equation 
\[
\left|a+B\right|\partial_{\theta}\Phi=P\Phi+\frac{B}{\left|a+B\right|}\partial_{x}\Phi-\left[1-\frac{B^{2}}{\left|a+B\right|}\right]\partial_{y}\Phi+\frac{1}{2}\frac{1}{\left|a+B\right|}\partial_{xx}\Phi+\frac{B}{\left|a+B\right|}\partial_{xy}\Phi-\frac{1}{2}\frac{aB}{\left|a+B\right|}\partial_{yy}\Phi-d\Phi|\Phi|^{2}
\]
(compare Eq. (2) from the main text) equipped with the boundary conditions
$\Phi(\theta,x,y)=e^{i\phi}\Phi(\theta,x-1,y)$ and $\Phi(\theta,x,y)=\exp\left[i\left(\xi-\phi\left[\frac{\kappa}{\varepsilon}\right]_{i}\right)\right]\Phi\left(\theta,x-\left\{ \frac{\kappa}{\varepsilon}\right\} _{f\mbox{ }},y-\kappa\right)$.
The simplest case of the periodic boundary conditions on the domain
$\left[0,1\right]^{2}$ arise for $\xi=\varphi=0$ (real parameters
$c$ and $b$), $\kappa=1$, and $\left\{ \frac{\kappa}{\varepsilon}\right\} _{f\mbox{ }}=0$.
The last condition means that the $\tau_{2}/\tau_{1}$ is a large
but integer number. Note that the proposed derivation is technically
different from the one given in \cite{Giacomelli1996} for the case
of one delay. Just to mention a few, the differences are in the way
how boundary conditions are handled and how different timescales appear
in the normal form equation. For instance, the timescale $\sim\tau^{3}$
appears for the one-delay case as the temporal variable of the normal
form, while here this is just a drifting part of the space coordinates
$x$ and $y$.

\begin{figure}
\includegraphics[width=0.5\linewidth]{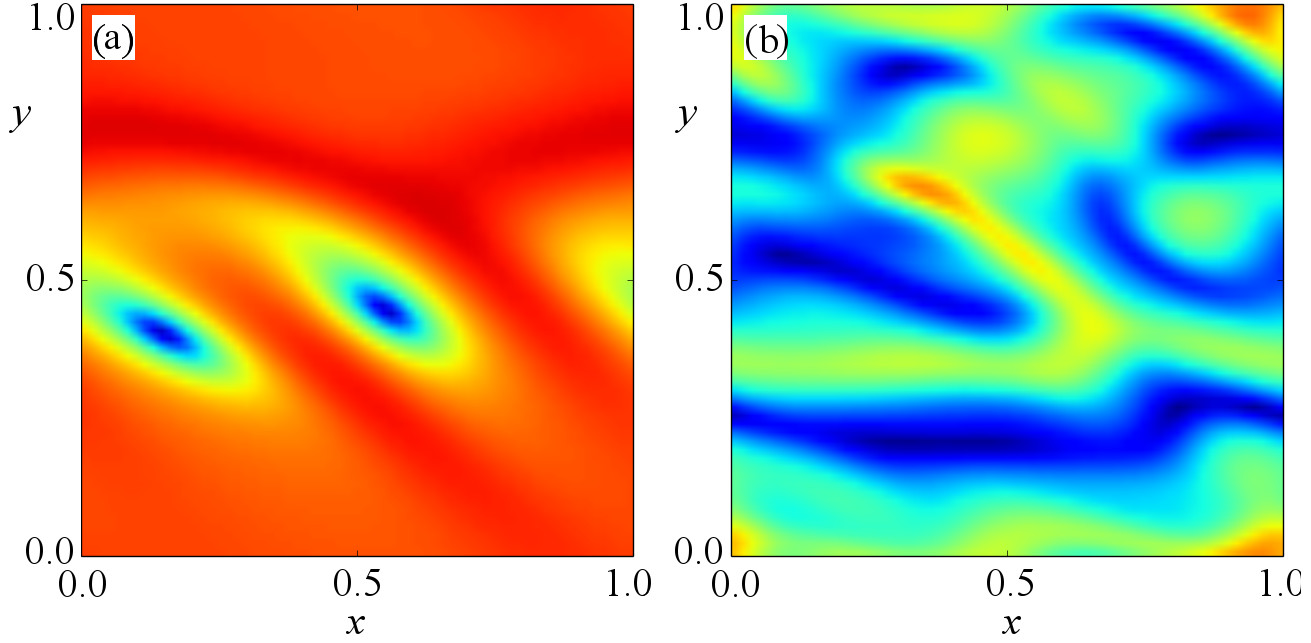}

\caption{Snapshots for the solutions of the Ginzburg-Landau normal form equation
(Eq. (2) in the main text). (a) Spiral defects, parameter values:
$p=250,$ $a_{1}=1.11$, $a_{2}=-1.22$, $a_{3}=1.39$, $a_{4}=1.11$,
$a_{5}=0.56$, $d=-0.75+i$. (b) Defect turbulence, parameter values
are the same except for $d=-0.1+i$. Initial values are random and
close to zero. \label{fig:Snapshots-for-the}}
\end{figure}

Figure \ref{fig:Snapshots-for-the} shows the snapshots of the solutions
of the normal form equation corresponding to the defects and turbulence
regimes (compare with Figs.~1 and 2 form the main text).

\bibliographystyle{apsrev}

\begin{thebibliography}{26}
\expandafter\ifx\csname natexlab\endcsname\relax\def\natexlab#1{#1}\fi
\expandafter\ifx\csname bibnamefont\endcsname\relax
  \def\bibnamefont#1{#1}\fi
\expandafter\ifx\csname bibfnamefont\endcsname\relax
  \def\bibfnamefont#1{#1}\fi
\expandafter\ifx\csname citenamefont\endcsname\relax
  \def\citenamefont#1{#1}\fi
\expandafter\ifx\csname url\endcsname\relax
  \def\url#1{\texttt{#1}}\fi
\expandafter\ifx\csname urlprefix\endcsname\relax\def\urlprefix{URL }\fi
\providecommand{\bibinfo}[2]{#2}
\providecommand{\eprint}[2][]{\url{#2}}

\bibitem[{\citenamefont{Li et~al.}(2011)\citenamefont{Li, Cohen, Murphy, and
  Roy}}]{Li2011}
\bibinfo{author}{\bibfnamefont{X.}~\bibnamefont{Li}},
  \bibinfo{author}{\bibfnamefont{A.~B.} \bibnamefont{Cohen}},
  \bibinfo{author}{\bibfnamefont{T.~E.} \bibnamefont{Murphy}},
  \bibnamefont{and} \bibinfo{author}{\bibfnamefont{R.}~\bibnamefont{Roy}},
  \bibinfo{journal}{Optics Lett.} \textbf{\bibinfo{volume}{36}},
  \bibinfo{pages}{1020} (\bibinfo{year}{2011}).

\bibitem[{\citenamefont{Zamora-Munt et~al.}(2010)\citenamefont{Zamora-Munt,
  Masoller, Garcia-Ojalvo, and Roy}}]{Zamora-Munt2010}
\bibinfo{author}{\bibfnamefont{J.}~\bibnamefont{Zamora-Munt}},
  \bibinfo{author}{\bibfnamefont{C.}~\bibnamefont{Masoller}},
  \bibinfo{author}{\bibfnamefont{J.}~\bibnamefont{Garcia-Ojalvo}},
  \bibnamefont{and} \bibinfo{author}{\bibfnamefont{R.}~\bibnamefont{Roy}},
  \bibinfo{journal}{Phys. Rev. Lett.} \textbf{\bibinfo{volume}{105}},
  \bibinfo{pages}{264101} (\bibinfo{year}{2010}).

\bibitem[{\citenamefont{Giacomelli et~al.}(2012)\citenamefont{Giacomelli,
  Marino, Zaks, and Yanchuk}}]{Giacomelli2012}
\bibinfo{author}{\bibfnamefont{G.}~\bibnamefont{Giacomelli}},
  \bibinfo{author}{\bibfnamefont{F.}~\bibnamefont{Marino}},
  \bibinfo{author}{\bibfnamefont{M.~A.} \bibnamefont{Zaks}}, \bibnamefont{and}
  \bibinfo{author}{\bibfnamefont{S.}~\bibnamefont{Yanchuk}},
  \bibinfo{journal}{EPL (Europhysics Letters)} \textbf{\bibinfo{volume}{99}},
  \bibinfo{pages}{58005} (\bibinfo{year}{2012}).

\bibitem[{\citenamefont{Soriano et~al.}(2013)\citenamefont{Soriano,
  Gar{c\'\i}a-Ojalvo, Mirasso, and Fischer}}]{Soriano2013}
\bibinfo{author}{\bibfnamefont{M.~C.} \bibnamefont{Soriano}},
  \bibinfo{author}{\bibfnamefont{J.}~\bibnamefont{Gar{c\'\i}a-Ojalvo}},
  \bibinfo{author}{\bibfnamefont{C.~R.} \bibnamefont{Mirasso}},
  \bibnamefont{and} \bibinfo{author}{\bibfnamefont{I.}~\bibnamefont{Fischer}},
  \bibinfo{journal}{Rev. Mod. Phys.} \textbf{\bibinfo{volume}{85}},
  \bibinfo{pages}{421} (\bibinfo{year}{2013}).

\bibitem[{\citenamefont{Szalai and Orosz}(2013)}]{Szalai2013}
\bibinfo{author}{\bibfnamefont{R.}~\bibnamefont{Szalai}} \bibnamefont{and}
  \bibinfo{author}{\bibfnamefont{G.}~\bibnamefont{Orosz}},
  \bibinfo{journal}{Phys. Rev. E} \textbf{\bibinfo{volume}{88}},
  \bibinfo{pages}{040902} (\bibinfo{year}{2013}).

\bibitem[{\citenamefont{Izhikevich}(2006)}]{Izhikevich2006}
\bibinfo{author}{\bibfnamefont{E.~M.} \bibnamefont{Izhikevich}},
  \bibinfo{journal}{Neural Computation} \textbf{\bibinfo{volume}{18}},
  \bibinfo{pages}{245} (\bibinfo{year}{2006}).

\bibitem[{\citenamefont{Appeltant et~al.}({2011})\citenamefont{Appeltant,
  Soriano, Van~der Sande, Danckaert, Massar, Dambre, Schrauwen, Mirasso, and
  Fischer}}]{Appeltant2011}
\bibinfo{author}{\bibfnamefont{L.}~\bibnamefont{Appeltant}},
  \bibinfo{author}{\bibfnamefont{M.~C.} \bibnamefont{Soriano}},
  \bibinfo{author}{\bibfnamefont{G.}~\bibnamefont{Van~der Sande}},
  \bibinfo{author}{\bibfnamefont{J.}~\bibnamefont{Danckaert}},
  \bibinfo{author}{\bibfnamefont{S.}~\bibnamefont{Massar}},
  \bibinfo{author}{\bibfnamefont{J.}~\bibnamefont{Dambre}},
  \bibinfo{author}{\bibfnamefont{B.}~\bibnamefont{Schrauwen}},
  \bibinfo{author}{\bibfnamefont{C.~R.} \bibnamefont{Mirasso}},
  \bibnamefont{and} \bibinfo{author}{\bibfnamefont{I.}~\bibnamefont{Fischer}},
  \bibinfo{journal}{Nature Comm} \textbf{\bibinfo{volume}{{2}}}
  (\bibinfo{year}{{2011}}).

\bibitem[{\citenamefont{Erneux}(2009)}]{Erneux2009}
\bibinfo{author}{\bibfnamefont{T.}~\bibnamefont{Erneux}},
  \emph{\bibinfo{title}{Applied Delay Differential Equations}}
  (\bibinfo{publisher}{Springer}, \bibinfo{year}{2009}).

\bibitem[{\citenamefont{Farmer}(1982)}]{Farmer1982}
\bibinfo{author}{\bibfnamefont{J.~D.} \bibnamefont{Farmer}},
  \bibinfo{journal}{Physica D} \textbf{\bibinfo{volume}{4}},
  \bibinfo{pages}{366} (\bibinfo{year}{1982}).

\bibitem[{\citenamefont{Giacomelli et~al.}(1995)\citenamefont{Giacomelli,
  Lepri, and Politi}}]{Giacomelli1995}
\bibinfo{author}{\bibfnamefont{G.}~\bibnamefont{Giacomelli}},
  \bibinfo{author}{\bibfnamefont{S.}~\bibnamefont{Lepri}}, \bibnamefont{and}
  \bibinfo{author}{\bibfnamefont{A.}~\bibnamefont{Politi}},
  \bibinfo{journal}{Phys. Rev. E} \textbf{\bibinfo{volume}{51}},
  \bibinfo{pages}{3939} (\bibinfo{year}{1995}).

\bibitem[{\citenamefont{Heiligenthal et~al.}(2011)\citenamefont{Heiligenthal,
  Dahms, Yanchuk, J\"ungling, Flunkert, Kanter, Sch\"oll, and
  Kinzel}}]{Heiligenthal2011}
\bibinfo{author}{\bibfnamefont{S.}~\bibnamefont{Heiligenthal}},
  \bibinfo{author}{\bibfnamefont{T.}~\bibnamefont{Dahms}},
  \bibinfo{author}{\bibfnamefont{S.}~\bibnamefont{Yanchuk}},
  \bibinfo{author}{\bibfnamefont{T.}~\bibnamefont{J\"ungling}},
  \bibinfo{author}{\bibfnamefont{V.}~\bibnamefont{Flunkert}},
  \bibinfo{author}{\bibfnamefont{I.}~\bibnamefont{Kanter}},
  \bibinfo{author}{\bibfnamefont{E.}~\bibnamefont{Sch\"oll}}, \bibnamefont{and}
  \bibinfo{author}{\bibfnamefont{W.}~\bibnamefont{Kinzel}},
  \bibinfo{journal}{Phys. Rev. Lett.} \textbf{\bibinfo{volume}{107}},
  \bibinfo{pages}{234102} (\bibinfo{year}{2011}).

\bibitem[{\citenamefont{D'Huys et~al.}(2013)\citenamefont{D'Huys, Zeeb,
  J{\"u}ngling, Heiligenthal, Yanchuk, and Kinzel}}]{DHuys2013}
\bibinfo{author}{\bibfnamefont{O.}~\bibnamefont{D'Huys}},
  \bibinfo{author}{\bibfnamefont{S.}~\bibnamefont{Zeeb}},
  \bibinfo{author}{\bibfnamefont{T.}~\bibnamefont{J{\"u}ngling}},
  \bibinfo{author}{\bibfnamefont{S.}~\bibnamefont{Heiligenthal}},
  \bibinfo{author}{\bibfnamefont{S.}~\bibnamefont{Yanchuk}}, \bibnamefont{and}
  \bibinfo{author}{\bibfnamefont{W.}~\bibnamefont{Kinzel}},
  \bibinfo{journal}{EPL (Europhysics Letters)} \textbf{\bibinfo{volume}{103}},
  \bibinfo{pages}{10013} (\bibinfo{year}{2013}).

\bibitem[{\citenamefont{Giacomelli and Politi}(1996)}]{Giacomelli1996}
\bibinfo{author}{\bibfnamefont{G.}~\bibnamefont{Giacomelli}} \bibnamefont{and}
  \bibinfo{author}{\bibfnamefont{A.}~\bibnamefont{Politi}},
  \bibinfo{journal}{Phys. Rev. Lett.} \textbf{\bibinfo{volume}{76}},
  \bibinfo{pages}{2686} (\bibinfo{year}{1996}).

\bibitem[{\citenamefont{Wolfrum and Yanchuk}(2006)}]{Wolfrum2006}
\bibinfo{author}{\bibfnamefont{M.}~\bibnamefont{Wolfrum}} \bibnamefont{and}
  \bibinfo{author}{\bibfnamefont{S.}~\bibnamefont{Yanchuk}},
  \bibinfo{journal}{Phys. Rev. Lett.} \textbf{\bibinfo{volume}{96}},
  \bibinfo{eid}{220201} (\bibinfo{year}{2006}).

\bibitem[{\citenamefont{Arecchi et~al.}(1992)\citenamefont{Arecchi, Giacomelli,
  Lapucci, and Meucci}}]{Arecchi1992}
\bibinfo{author}{\bibfnamefont{F.}~\bibnamefont{Arecchi}},
  \bibinfo{author}{\bibfnamefont{G.}~\bibnamefont{Giacomelli}},
  \bibinfo{author}{\bibfnamefont{A.}~\bibnamefont{Lapucci}}, \bibnamefont{and}
  \bibinfo{author}{\bibfnamefont{R.}~\bibnamefont{Meucci}},
  \bibinfo{journal}{Phys. Rev. A} \textbf{\bibinfo{volume}{45}},
  \bibinfo{pages}{4225} (\bibinfo{year}{1992}).

\bibitem[{\citenamefont{Giacomelli et~al.}(1994)\citenamefont{Giacomelli,
  Meucci, Politi, and Arecchi}}]{Giacomelli1994}
\bibinfo{author}{\bibfnamefont{G.}~\bibnamefont{Giacomelli}},
  \bibinfo{author}{\bibfnamefont{R.}~\bibnamefont{Meucci}},
  \bibinfo{author}{\bibfnamefont{A.}~\bibnamefont{Politi}}, \bibnamefont{and}
  \bibinfo{author}{\bibfnamefont{F.~T.} \bibnamefont{Arecchi}},
  \bibinfo{journal}{Phys. Rev. Lett.} \textbf{\bibinfo{volume}{73}},
  \bibinfo{pages}{1099} (\bibinfo{year}{1994}).

\bibitem[{\citenamefont{Larger et~al.}(2013)\citenamefont{Larger, Penkovsky,
  and Maistrenko}}]{Larger2013}
\bibinfo{author}{\bibfnamefont{L.}~\bibnamefont{Larger}},
  \bibinfo{author}{\bibfnamefont{B.}~\bibnamefont{Penkovsky}},
  \bibnamefont{and}
  \bibinfo{author}{\bibfnamefont{Y.}~\bibnamefont{Maistrenko}},
  \bibinfo{journal}{Phys. Rev. Lett.} \textbf{\bibinfo{volume}{111}},
  \bibinfo{pages}{054103} (\bibinfo{year}{2013}).

\bibitem[{\citenamefont{Chat{\'e} and Manneville}(1996)}]{Chate1996}
\bibinfo{author}{\bibfnamefont{H.}~\bibnamefont{Chat{\'e}}} \bibnamefont{and}
  \bibinfo{author}{\bibfnamefont{P.}~\bibnamefont{Manneville}},
  \bibinfo{journal}{Physica A: Statistical Mechanics and its Applications}
  \textbf{\bibinfo{volume}{224}}, \bibinfo{pages}{348} (\bibinfo{year}{1996}).

\bibitem[{\citenamefont{Aranson and Kramer}(2002)}]{Aranson2002}
\bibinfo{author}{\bibfnamefont{I.}~\bibnamefont{Aranson}} \bibnamefont{and}
  \bibinfo{author}{\bibfnamefont{L.}~\bibnamefont{Kramer}},
  \bibinfo{journal}{Rev. Mod. Phys.} \textbf{\bibinfo{volume}{74}},
  \bibinfo{pages}{99} (\bibinfo{year}{2002}).

\bibitem[{\citenamefont{Deissler and Kaneko}(1987)}]{Deissler1987}
\bibinfo{author}{\bibfnamefont{R.~J.} \bibnamefont{Deissler}} \bibnamefont{and}
  \bibinfo{author}{\bibfnamefont{K.}~\bibnamefont{Kaneko}},
  \bibinfo{journal}{Phys. Lett. A} \textbf{\bibinfo{volume}{119}},
  \bibinfo{pages}{397} (\bibinfo{year}{1987}).

\bibitem[{\citenamefont{Lang and Kobayashi}(1980)}]{Lang1980}
\bibinfo{author}{\bibfnamefont{R.}~\bibnamefont{Lang}} \bibnamefont{and}
  \bibinfo{author}{\bibfnamefont{K.}~\bibnamefont{Kobayashi}},
  \bibinfo{journal}{IEEE J. Quantum Electron.} \textbf{\bibinfo{volume}{16}},
  \bibinfo{pages}{347} (\bibinfo{year}{1980}).

\bibitem[{\citenamefont{Barland et~al.}(2005)\citenamefont{Barland, Spinicelli,
  Giacomelli, and Marin}}]{Barland2005}
\bibinfo{author}{\bibfnamefont{S.}~\bibnamefont{Barland}},
  \bibinfo{author}{\bibfnamefont{P.}~\bibnamefont{Spinicelli}},
  \bibinfo{author}{\bibfnamefont{G.}~\bibnamefont{Giacomelli}},
  \bibnamefont{and} \bibinfo{author}{\bibfnamefont{F.}~\bibnamefont{Marin}},
  \bibinfo{journal}{IEEE J. Quantum Electron.} \textbf{\bibinfo{volume}{41}},
  \bibinfo{pages}{1235} (\bibinfo{year}{2005}).

\bibitem[{\citenamefont{Wolfrum et~al.}(2010)\citenamefont{Wolfrum, Yanchuk,
  {H\"ovel}, and {Sch\"oll}}}]{Wolfrum2010}
\bibinfo{author}{\bibfnamefont{M.}~\bibnamefont{Wolfrum}},
  \bibinfo{author}{\bibfnamefont{S.}~\bibnamefont{Yanchuk}},
  \bibinfo{author}{\bibfnamefont{P.}~\bibnamefont{{H\"ovel}}},
  \bibnamefont{and}
  \bibinfo{author}{\bibfnamefont{E.}~\bibnamefont{{Sch\"oll}}},
  \bibinfo{journal}{Eur. Phys. J. Special Topics}
  \textbf{\bibinfo{volume}{191}}, \bibinfo{pages}{91} (\bibinfo{year}{2010}).

\bibitem[{\citenamefont{Lepri et~al.}(1994)\citenamefont{Lepri, Giacomelli,
  Politi, and Arecchi}}]{Lepri1993}
\bibinfo{author}{\bibfnamefont{S.}~\bibnamefont{Lepri}},
  \bibinfo{author}{\bibfnamefont{G.}~\bibnamefont{Giacomelli}},
  \bibinfo{author}{\bibfnamefont{A.}~\bibnamefont{Politi}}, \bibnamefont{and}
  \bibinfo{author}{\bibfnamefont{F.~T.} \bibnamefont{Arecchi}},
  \bibinfo{journal}{Physica D} \textbf{\bibinfo{volume}{70}},
  \bibinfo{pages}{235} (\bibinfo{year}{1994}).

\bibitem[{\citenamefont{Yanchuk and Wolfrum}(2010)}]{Yanchuk2010a}
\bibinfo{author}{\bibfnamefont{S.}~\bibnamefont{Yanchuk}} \bibnamefont{and}
  \bibinfo{author}{\bibfnamefont{M.}~\bibnamefont{Wolfrum}},
  \bibinfo{journal}{SIAM J Appl Dyn Syst} \textbf{\bibinfo{volume}{9}},
  \bibinfo{pages}{519} (\bibinfo{year}{2010}).

\bibitem[{\citenamefont{Lichtner et~al.}(2011)\citenamefont{Lichtner, Wolfrum,
  and Yanchuk}}]{Lichtner2011}
\bibinfo{author}{\bibfnamefont{M.}~\bibnamefont{Lichtner}},
  \bibinfo{author}{\bibfnamefont{M.}~\bibnamefont{Wolfrum}}, \bibnamefont{and}
  \bibinfo{author}{\bibfnamefont{S.}~\bibnamefont{Yanchuk}},
  \bibinfo{journal}{SIAM J. Math. Anal.} \textbf{\bibinfo{volume}{43}},
  \bibinfo{pages}{788} (\bibinfo{year}{2011}).

\end{thebibliography}

\end{document}